\documentclass[lettersize,journal]{IEEEtran}
\usepackage{amsmath,amsfonts}
\usepackage{algorithmic}
\usepackage{array}
\usepackage[caption=false,font=normalsize,labelfont=sf,textfont=sf]{subfig}
\usepackage{textcomp}
\usepackage{stfloats}
\usepackage{url}
\usepackage{verbatim}
\usepackage{graphicx}
\usepackage{hyperref}
\usepackage{academicons}
\usepackage{xcolor}
\usepackage{upgreek}
\usepackage{siunitx}
\usepackage{multirow}
\usepackage{booktabs}
\usepackage{comment}    
\usepackage[inline]{enumitem}   
\def\BibTeX{{\rm B\kern-.05em{\sc i\kern-.025em b}\kern-.08em
    T\kern-.1667em\lower.7ex\hbox{E}\kern-.125emX}}
\usepackage{balance}

\usepackage{scalerel}
\usepackage{tikz}
\usetikzlibrary{svg.path}

\definecolor{orcidlogocol}{HTML}{A6CE39}
\tikzset{
  orcidlogo/.pic={
    \fill[orcidlogocol] svg{M256,128c0,70.7-57.3,128-128,128C57.3,256,0,198.7,0,128C0,57.3,57.3,0,128,0C198.7,0,256,57.3,256,128z};
    \fill[white] svg{M86.3,186.2H70.9V79.1h15.4v48.4V186.2z}
                 svg{M108.9,79.1h41.6c39.6,0,57,28.3,57,53.6c0,27.5-21.5,53.6-56.8,53.6h-41.8V79.1z M124.3,172.4h24.5c34.9,0,42.9-26.5,42.9-39.7c0-21.5-13.7-39.7-43.7-39.7h-23.7V172.4z}
                 svg{M88.7,56.8c0,5.5-4.5,10.1-10.1,10.1c-5.6,0-10.1-4.6-10.1-10.1c0-5.6,4.5-10.1,10.1-10.1C84.2,46.7,88.7,51.3,88.7,56.8z};
  }
}

\newcommand\orcidicon[1]{\href{https://orcid.org/#1}{\mbox{\scalerel*{
\begin{tikzpicture}[yscale=-1,transform shape]
\pic{orcidlogo};
\end{tikzpicture}
}{|}}}}

\usepackage{hyperref} 

\begin{document}
\title{A Survey of Spiking Neural Network Accelerator on FPGA}
\author{

    Murat Isik  \orcidicon{0000-0002-0907-7253}

\thanks{Manuscript created June 2023; This work was developed by the IEEE Publication Technology Department. This work is distributed under the \LaTeX \ Project Public License (LPPL) ( http://www.latex-project.org/ ) version 1.3. A copy of the LPPL, version 1.3, is included in the base \LaTeX \ documentation of all distributions of \LaTeX \ released 2003/12/01 or later. The opinions expressed here are entirely that of the author. No warranty is expressed or implied. User assumes all risk.}
\thanks{Murat Isik is with Stanford University, 94305 Stanford, CA, United States of America \href{mailto:mrtisik@stanford.edu}{mrtisik@stanford.edu}}

}
\IEEEoverridecommandlockouts

\markboth{Journal of \LaTeX\ Class Files,~Vol.~18, No.~9, June~2023}%
{How to Use the IEEEtran \LaTeX \ Templates}

\maketitle
\begin{abstract}

Due to the ability to implement customized topology, FPGA is increasingly used to deploy SNNs in both embedded and high-performance applications. In this paper, we survey state-of-the-art SNN implementations and their applications on FPGA. We collect the recent widely-used spiking neuron models, network structures, and signal encoding formats, followed by the enumeration of related hardware design schemes for FPGA-based SNN implementations. Compared with the previous surveys, this manuscript enumerates the application instances that applied the above-mentioned technical schemes in recent research. Based on that, we discuss the actual acceleration potential of implementing SNN on FPGA. According to our above discussion, the upcoming trends are discussed in this paper and give a guideline for further advancement in related subjects.
\end{abstract}

\begin{IEEEkeywords}
Neural Networks, Spiking Neural Network, FPGA Implementation, Hardware Accelerator, Survey.
\end{IEEEkeywords}

\section{Introduction}

\IEEEPARstart{S}{\emph{piking}}~\emph{Neural Network} (SNN) has been one of the most extensive research subjects in recent decades. Researchers successfully deployed the related instances in various application scenarios, such as speech recognition, biomedical analysis, and self-driving cars. SNNs are inspired by the biological neural system with the understanding of brain functionalities. The related works are more biologically plausible in both information transmission across neurons and internal neuron signal processing. Therefore, this technique causes a paradigm shift in the field of neural network research. The goal of SNNs is to develop a computational system modeling the behavior of real neurons. However, the complexities of the network model and corresponding computation requirements in SNN inference are rapidly increasing. The trade-off between hardware and power consumption and acceleration performance has become a research topic of importance. This leads to an actual requirement for customized hardware accelerators to achieve higher computing/power efficiency, especially for embedded and lightweight applications. Recent research shows that SNNs have the following excellent hardware implementation features: SNNs communicate across neurons using spikes, which are equivalent to a single bit in terms of logic resources and decrease the logic occupation. Therefore,~\emph{Field Programmable Gate Arrays} (FPGAs) become a potential hardware platform for SNN acceleration, based on its features of logic reconfigurability, high power efficiency, and support of computing parallelization and bit-level operations.

Recent research has deployed SNNs on different platforms, such as~\emph{Central Processing Units} (CPUs),~\emph{Graphics Processing Units} (GPUs) and~\emph{Application-Specific Integrated Circuits} (ASICs). However, due to the restricted memory bandwidth, SNNs on CPU/GPU consume high power overhead with limited throughput~\cite{huynh2022implementing, isik2023design}. To achieve the best performance and energy efficiency, many researchers have focused on building custom ASICs for accelerating network inference workloads. Despite being an attractive solution, ASICs cannot offer sufficient flexibility to accommodate the rapid evolution of~\emph{Neural Network} (NN) models. The emergence of new types of layers within NN, including branching, elementwise addition, and batch normalization layers, has been seen in more recent models that require flexibility. Further, the high~\emph{Non-Recurring Engineering} (NRE) cost and time for design, verification, and fabrication of a large ASIC chip make it difficult to keep pace with the rapid model improvements in this space~\cite{boutros2018you}. Consequently, FPGAs serve as configurable tools, facilitating the creation of unique logic, which may mitigate the limitations on the execution of neural networks. As an outcome, due to the apparent benefits of the SNN on hardware implementation, one of the current research hotspots is the development of hardware systems supporting SNN inference based on FPGA to achieve high throughput and power efficiency \cite{isik2022design}.

\subsection{Contributions of Our Survey}

In this manuscript, we survey the upcoming techniques of FPGA-based SNN accelerators and its application instance in various scenarios. The major contributions of our survey are as follows:

\begin{itemize}
    \item We categorize the widely-used spiking neuron models, spiking signal format, currently explored SNNs, SNN training tools, etc.
    
    \item Based on the above-mentioned background, we survey the feasible techniques for hardware design of SNN accelerator on FPGA and analyze the difficulties in the implementation, including accelerator architectures, optimization, and upcoming solutions. 
    
    \item We enumerated the related application-based SNN acceleration on FPGA. 
    
    \item We examine the research trends and optimization directions in developing FPGA-based SNN accelerators based on the above works.
\end{itemize}

\subsection{Organization}
This manuscript is structured as following \textbf{Section II} introduces the technology background of~\emph{Spiking Neural Networks}. \textbf{Section III} discusses the popular implementations and solutions for SNN accelerators on FPGA in recent works. \textbf{Section IV} briefly emulates the previous FPGA applications in different scenarios based on SNN. \textbf{Section V} analyzes the trend of further work and upcoming research for the SNN accelerator on FPGA. \textbf{Section VI} concludes the survey.

\section{Background}\label{sec:background}
\subsection{Coding Formats of Spiking Signals}

In recent artificial intelligence research, spiking neural networks have become a state-of-the-art topic for further modeling the working principle of bio-neural systems. Therefore, the input data need to be encoded into a spiking signal format. For instance,~\autoref{spiking_signal_fig} shows five most popular spiking coding formats in recent research~\cite{guo2021neural, von2017artificial}:

\begin{figure}[h]
    \centering\includegraphics[width=\columnwidth]{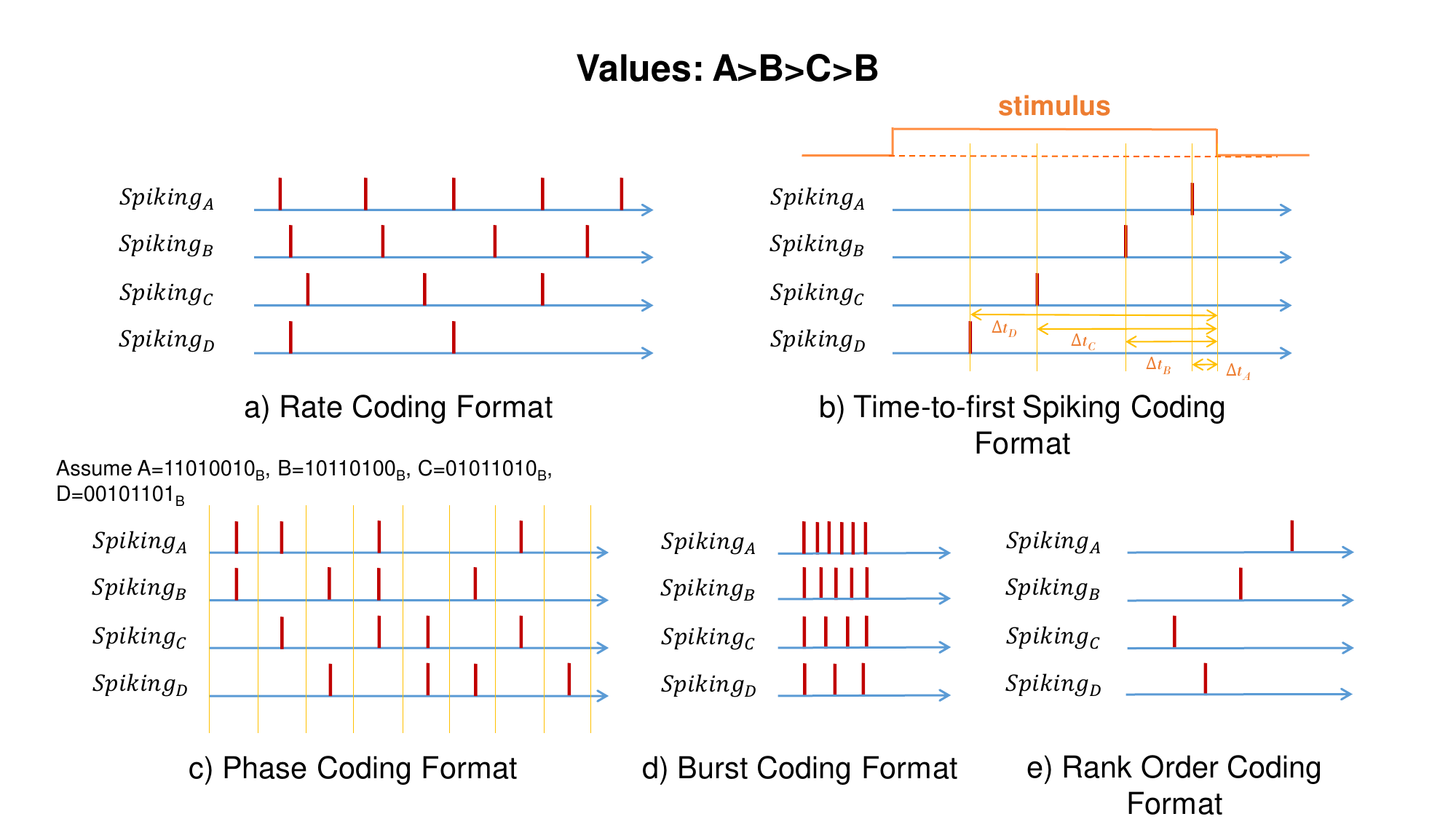}
    \centering\caption{Five popular spiking coding formats}
    \label{spiking_signal_fig}
\end{figure}

\begin{itemize}
    \item \emph{\textbf{Rate Coding Format: }} Rate coding is the most widely-used format in related works. This format converts the input value to a Poisson spiking train with the corresponding fire rate. As shown in~\autoref{spiking_signal_fig}a, larger input can be represented as the spiking train with a higher fire rate. 
    
    \item \emph{\textbf{Time-to-first Spiking Coding Format: }}~\emph{Time-to-first Spiking Coding} is a kind of temporal coding method. Instead of using fire rate to represent different inputs value, Time-to-first coding applies the fire latency in a stimulus duration to convert data as a spike. As shown in~\autoref{spiking_signal_fig}b, the spike of larger input will be fired earlier. Compared with the rate coding format, time-to-first coding only requires a single spike, data information will be encoded in the temporal latency.
    
    \item \emph{\textbf{Phase Coding Format: }} Above two coding methods need a conversion between data value and fire rate/latency. Different conversion solutions will influence the encoded spike trains. As shown in~\autoref{spiking_signal_fig}c, work~\cite{kim2018deep} explored another spiking encoding method that directly converts the binary format of input value as the spike train. The weight spike in this work will represent the significance of each input bit.
    
    \item \emph{\textbf{Burst Coding Format: }} To reduce the transmission duration of spiking signal, work~\cite{izhikevich2003bursts}\cite{eyherabide2009bursts} explored a burst transmission encoding method as shown in~\autoref{spiking_signal_fig}d. This method encodes the input value as the number of spikes and the inter-spike interval (ISI) in the burst. More intensive spikes in burst represent a larger input.
    
    \item \emph{\textbf{Rank Order Coding Format: }}~\autoref{spiking_signal_fig}d shows the scheme of~\emph{Rank Order Coding}~\cite{thorpe1998rank}. Instead of applying the spikes latency in stimulus to encode the inputs like~\emph{Time-to-first Spiking Coding} scheme, this encoding method encodes the input information as the spikes fire order ignoring the fire latency. Therefore, for $N$ pre-neurons, their outputs can be encoded as $2^N$ different data.
\end{itemize}

\subsection{Spiking Neuron Modelling and the Differences with Classic ANN Neuron}

\subsubsection{Spiking Neuron Modelling}
A bio-inspired artificial intelligence technology called SNN has become a state-of-the-art technology in recent years. Therefore, the implementations of spiking neurons also mimic the actual natural neural cells. Based on the cell structure modelling, there are three widely-explored spiking neuron models on FPGA accelerator implementations: i)~\emph{Hodgkin-Huxley (HH) Model}~\cite{hodgkin1952quantitative}, ii)~\emph{Izhikevich Model}~\cite{izhikevich2003simple}, iii)~\emph{Leaky Integrate-and-Fire Model} (LIF)~\cite{Abbott1999LapicquesIO}\cite{andrew2003spiking}.

\begin{itemize}
    \item \emph{\textbf{Hodgkin-Huxley (HH) Model: }} This model simplified the neural cell structure as a RC circuit model as shown in~\autoref{hh_model_fig}.
    
    \begin{figure}[h]
        \centerline{\includegraphics[width=\columnwidth]{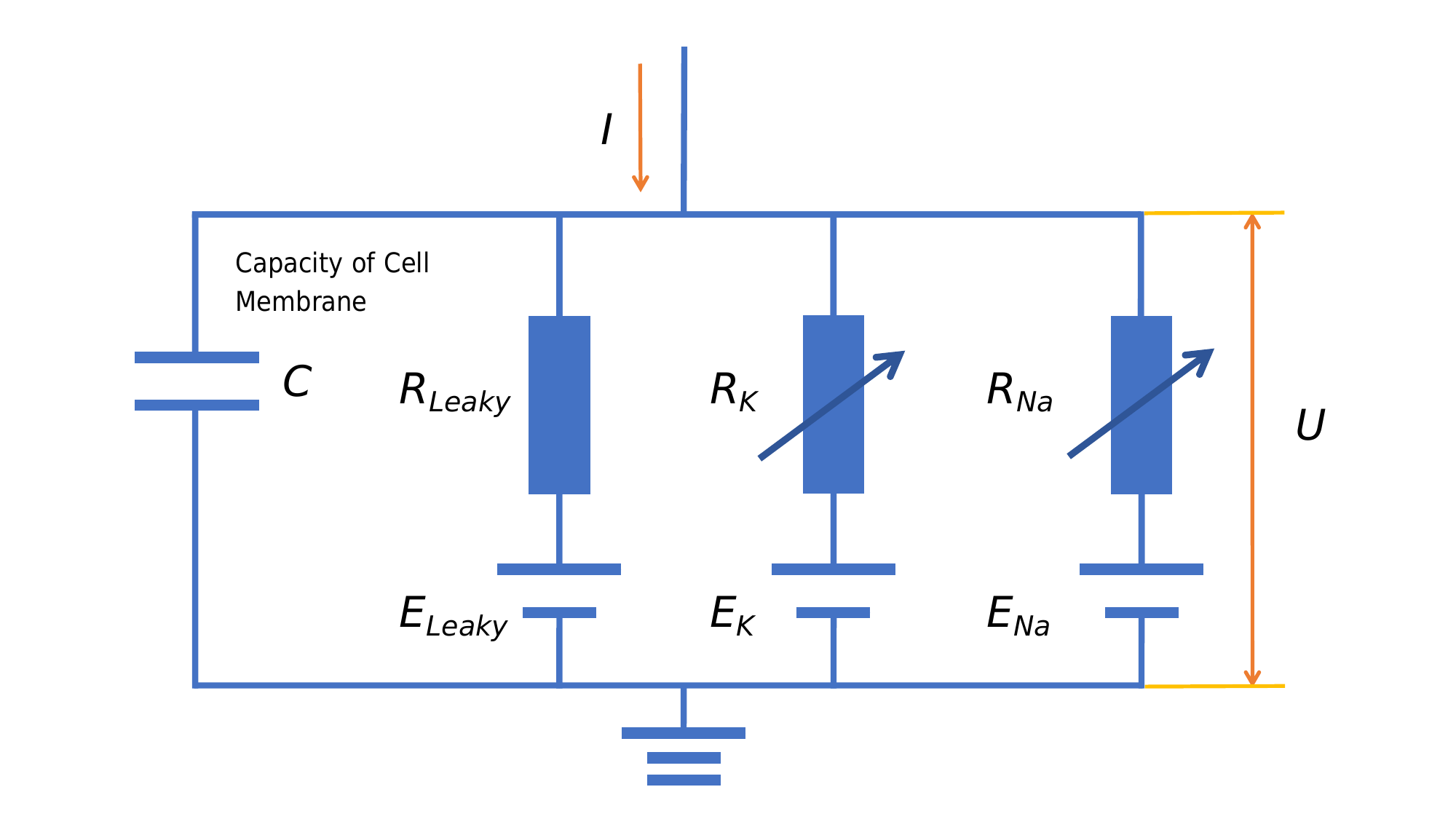}}
        \caption{RC Circuit Modelling of Hodgkin-Huxley Spiking Neuron Model}
        \label{hh_model_fig}
    \end{figure}
    
    In this model, $C$ represents the capacity of the cell membrane. $R_K$ and $R_{Na}$ mimic the sodium and potassium ion channels in neurons. $R_{Leaky}$ simulates the leaky channel to lose the charge in the membrane. Therefore,~\autoref{hh_formular}
    
    \begin{equation}
        \begin{aligned}
            \begin{aligned}
                I\left ( t \right ) & = I_{C}\left ( t \right ) + I_{K}\left ( t \right ) + I_{Na}\left ( t \right ) + I_{Leaky}\left ( t \right ) 
            \end{aligned}\\
            \begin{aligned}
                \Rightarrow C\frac{\mathrm{d} U}{\mathrm{d} t} & = I\left ( t \right ) - I_{K}\left ( t \right ) - I_{Na}\left ( t \right ) - I_{Leaky}\left ( t \right )
            \end{aligned}
        \end{aligned}
        \label{hh_formular}
    \end{equation}
    
    Because the conductance of ion channels in this model is simulated as the function of time and voltage,~\autoref{hh_formular} will be rewritten as:
    
    \begin{equation}
        \begin{aligned}
            & \begin{aligned}
                I\left ( t \right ) & = C\frac{\mathrm{d} U}{\mathrm{d} t} + g_{K}n\left ( t, U \right )^4 \cdot (U-U_{K}) \\
                & + g_{Na}m\left ( t, U \right )^3h\left ( t, U \right ) \cdot (U-U_{Na})\\ 
                & + g_{Leaky}\left ( t \right ) \cdot (U-U_{K})
            \end{aligned} \\
            & \begin{aligned}
                \left\{\begin{matrix}
                    \begin{aligned}
                        &\frac{\mathrm{d} n\left ( t, U \right ) }{\mathrm{d} t} = \alpha_{n} \left ( U \right )\cdot (1-n\left ( t, U \right )) - \beta_{n} \left ( U \right ) \cdot n\left ( t, U \right )\\
                        &\frac{\mathrm{d} m\left ( t, U \right ) }{\mathrm{d} t} = \alpha_{m} \left ( U \right )\cdot (1-m\left ( t, U \right )) - \beta_{m} \left ( U \right ) \cdot m\left ( t, U \right )\\
                        &\frac{\mathrm{d} h\left ( t, U \right ) }{\mathrm{d} t} = \alpha_{h} \left ( U \right )\cdot (1-h\left ( t, U \right )) - \beta_{h} \left ( U \right ) \cdot h\left ( t, U \right )
                    \end{aligned}
                \end{matrix}\right.
            \end{aligned} \\
            & \begin{aligned}
                \left\{\begin{matrix}
                    \begin{aligned}
                        \alpha_{i} \left ( U \right ) & = p_\infty \left ( U \right ) / \tau_p\\
                        \beta_{i} \left ( U \right ) & = (1 - p_\infty \left ( U \right )) / \tau_p 
                    \end{aligned}
                \end{matrix}\right.
            \end{aligned} 
        \end{aligned}
        \label{hh_formular_complete}
    \end{equation}

    In~\autoref{hh_formular_complete}, $n$, $m$, and $h$ are the conductance parameters for sodium and potassium ion channels, $p$ represent the steady state values for activation.
    \item \emph{\textbf{Izhikevich Model: }} Considering the parameters and complex computing in~\autoref{hh_formular_complete}, work~\cite{izhikevich2003simple} simplified the HH model based on bifurcation methodologies as~\autoref{izh_formular} to reduce the number of parameters. 
    
    \begin{equation}
        \begin{aligned}
            & \begin{aligned}
                v' & = 0.04v^2+5v+140-u+I\\
                u' & = a(bv-u)
            \end{aligned}\\
            & \begin{aligned}
                 \begin{matrix}
                if & v\geq 30mV, & then & 
                \left\{\begin{matrix}
                    \begin{aligned}
                        v & \leftarrow c\\ 
                        u & \leftarrow u+ d
                    \end{aligned}
                \end{matrix}\right.
                \end{matrix}
            \end{aligned}
        \end{aligned}
        \label{izh_formular}
    \end{equation}
    
    In~\autoref{izh_formular}, $v$ implies the membrane potential (input). $u$ means recover parameter. $a$ means how fast the membrane potential recovers, the typical value is 0.02. $b$ means the responsibility between $u$ and input $v$, typical value is 0.2. $c$ means the reset potential after spiking, the typical value is $-65mV$. $d$ means the increase value of $u$ after reset, typical value is $2mV$~\cite{Fang2022Izh}.
 
    \item \emph{\textbf{Leaky Integrate-and-Fire Model: }} 
    
    \begin{figure}[h]
        \centerline{\includegraphics[width=\columnwidth]{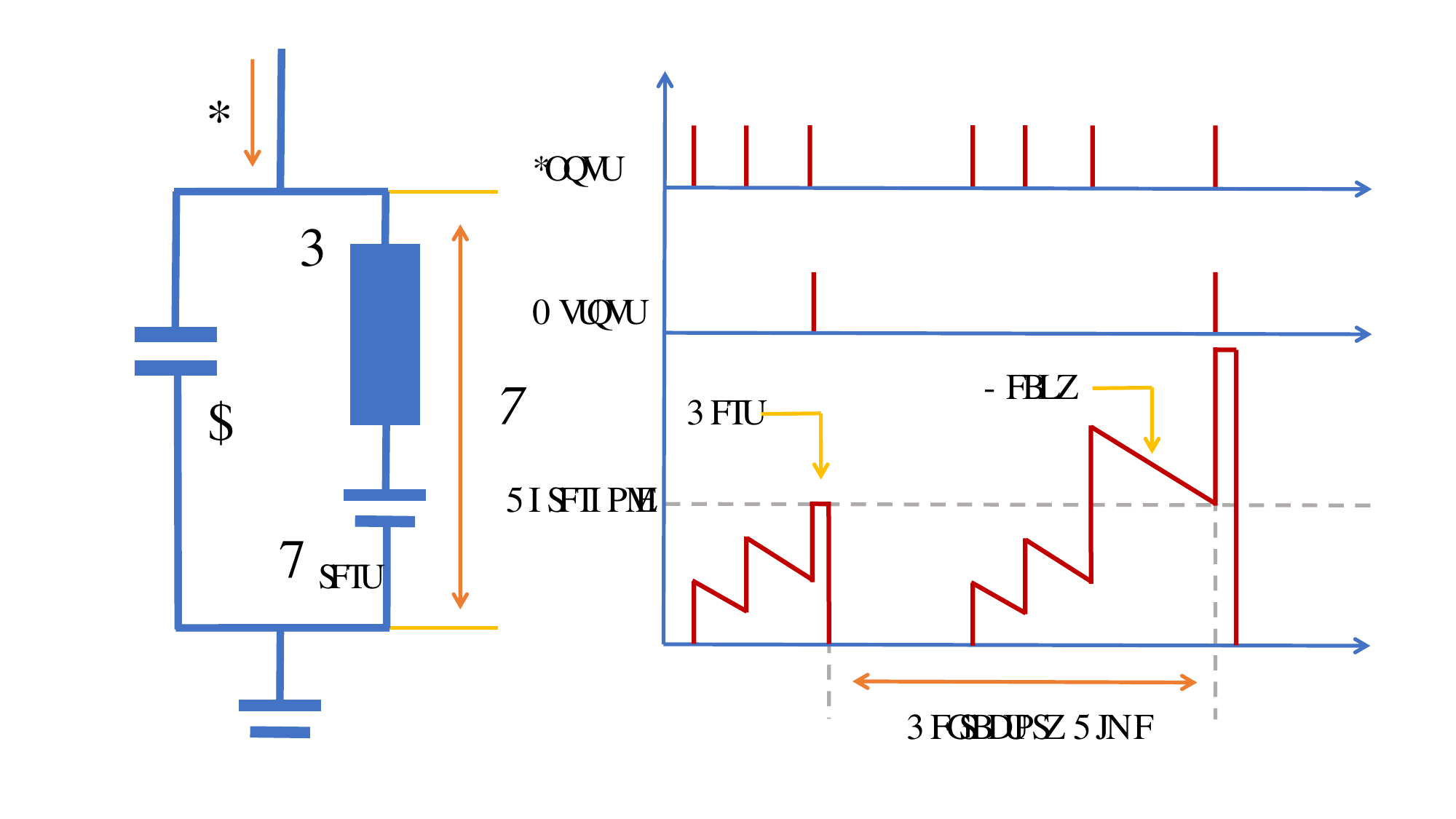}}
        \caption{RC Circuit Modelling of Leaky Integrate-and-Fire Spiking Neuron Model}
        \label{lif_model_fig}
    \end{figure}
    
    To further simplify the spiking neuron model, authors in~\cite{Abbott1999LapicquesIO, andrew2003spiking} discussed the simplification of the HH model as~\emph{Leaky Integrate-and-Fire} (LIF) model, which has been the most widely-used model in recent research. As shown in~\autoref{lif_model_fig}, LIF further simplified the HH model, the RC modeling formula is~\autoref{lif_formular}:
    
    \begin{equation}
        \begin{aligned}
            C\frac{\mathrm{d} U}{\mathrm{d} t} & = I\left ( t \right ) - \frac{V-V_{rest}}{R}
        \end{aligned}
        \label{lif_formular}
    \end{equation}
    
\end{itemize}

\subsubsection{Brief introduction of SNN Neurons hardware implementation}

~\autoref{spik_neuron_fig} shows the structure of neuron processing elements in SNNs hardware implementation. For SNN, when networks apply different neuron models and signal encoding schemes, the structures of neuron processing elements will be different. For instance, in the implementations based on LIF model and rate coding schemes, the neurons can consist of the following submodules: 1)~\emph{Multiplier}, 2)~\emph{Accumulator}, 3)~\emph{Thresholds}, and 4)~\emph{Spiking Encoder}:

\begin{figure}[h]
    \centerline{\includegraphics[width=\columnwidth]{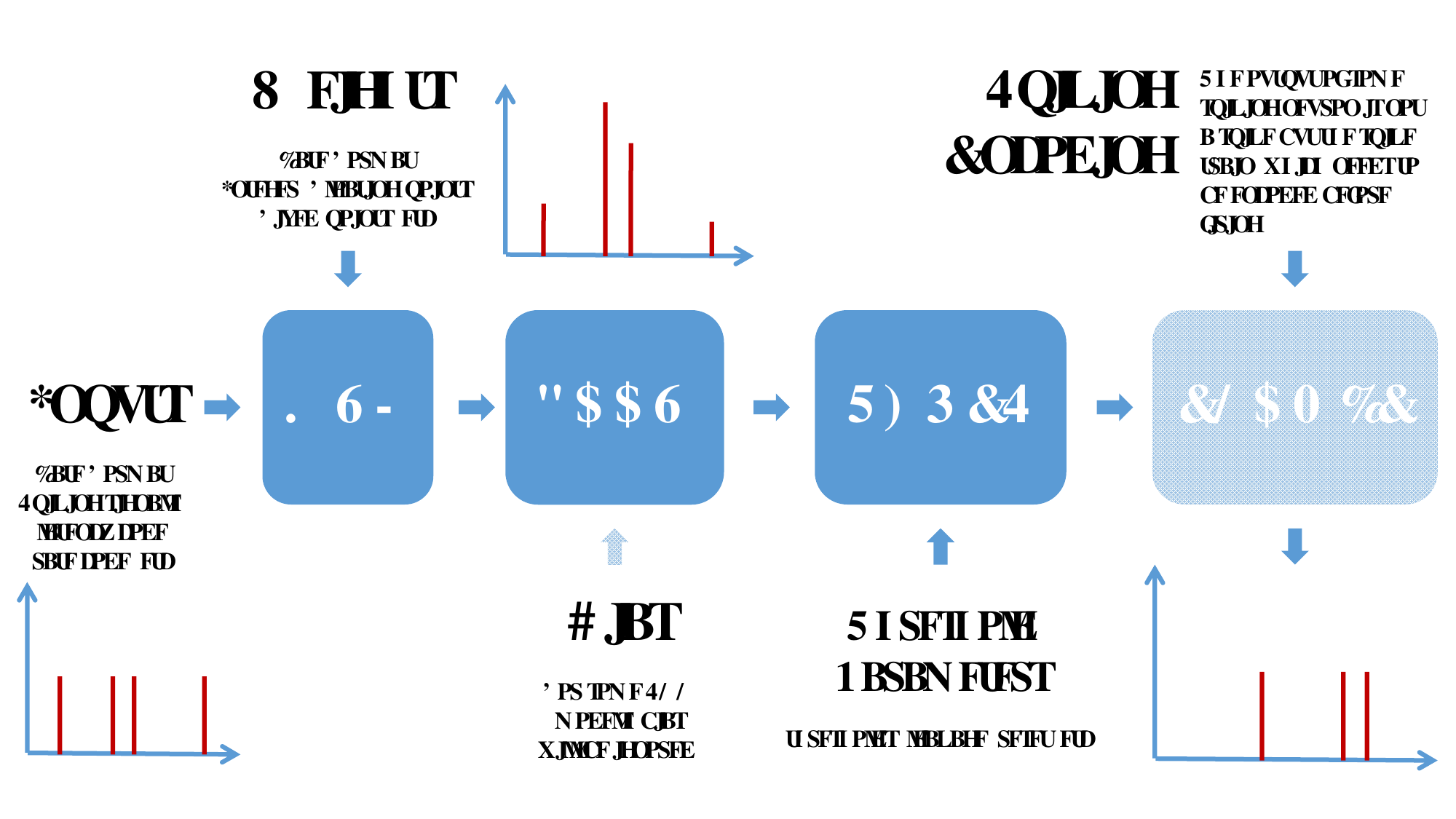}}
    \caption{Neuron Structure of Spiking Neural Networks}
    \label{spik_neuron_fig}
\end{figure}

\begin{itemize}
    \item \emph{\textbf{Multiplier}} and~\emph{\textbf{Accumulator}} compute the product sum of weighted spikes based on different data formats. However, storing network weights consumes a large number of on-chip memory resource on FPGA. Considering the limitation of hardware resources, in addition to the common floating-point and fixed-point formats, some low-precision data formats are widely explored in recent research, such as i)~\emph{Quantization}, ii)~\emph{Binarization}, iii)~\emph{Approximation Computation}, and iv)~\emph{Posit-floating Computation}:
    
    The authors of~\emph{Q-SpiNN}~\cite{putra2021q} and work~\cite{quan2020snn} implemented a~\emph{Quantized Spiking Neural Network} (QSNN) framework to reduce the memory consumption of network weights based on the low-precision integer scheme. Moreover, work~\emph{BS4NN}~\cite{kheradpisheh2021bs4nn} and work~\cite{wei2021binarized} further reduced the weight precision to the~\emph{Binarized Spiking Neural Network} (BSNN). Work~\emph{AxSNN}~\cite{sen2017approximate} explored the application of approximation computation on SNN. The authors of~\cite{silvaposit} researched the possibility of applying~\emph{Posit}-floating computation~\cite{posit1, posit2} on FPGA-based SNN accelerators.
    
    \item \emph{\textbf{Thresholds and Spiking Encoding}}: As shown in~\autoref{spik_neuron_fig}, the parameters of thresholds will be the leaky rate, refractory time, fire thresholds, etc. For some coding schemes, such as~\emph{Phase Coding}, the encoding module will be necessary. More technique details about the spiking neuron implementation will be discussed in Section III.
\end{itemize}

\subsection{SNN Model Training and Conversion Tools}
The simulation and evaluation of SNN accelerator implementations on FPGA require trained models. However, because of the difference between SNN and classic ANN, SNN training is still a challenge in recent research and limits the application of SNNs. Some upcoming toolkits support the SNN model generation-based conversion from trained ANN models or directly SNN training.

\subsubsection{ANN Model Conversion}

Because of the difficulties in synaptic weight learning on SNN training, some works explored the solution that converts an ANN model to SNN.~\emph{Spiking Neural Network Conversion Toolbox}~\cite{Rueckauer2017snn} implemented one python-based conversion tool from ANN to SNN supporting various ANN training libraries, such as \emph{Keras, PyTorch, Caffe,} etc. It also supports the SNN model export for evaluation based on SNN simulation tools, like \emph{pyNN}~\cite{pyNN}, \emph{Brain2}~\cite{Brain2}, \emph{sPyNNaker}~\cite{sPyNNaker}, etc.

\subsubsection{SNN Model Training} Compared with the ANN conversion, direct training of SNN can achieve higher accuracy. Some researchers explored the related SNN training algorithms~\cite{lee2016training, wade2010swat, wu2019direct}. Based on previous works, snnTorch~\cite{eshraghian2021training} implemented a PyTorch-based SNN training acceleration library. This library supports the training of~\emph{Spiking Convolution Neural Network} (SCNN) and~\emph{Spiking Long-Short Term Memory Network} (SLSTM) based on the LIF neuron model and rate/latency coding. The training based on this work can also be accelerated by applying the GPU platform through PyTorch.

\section{Techniques}


\subsection{Topology Conversion}
Spiking neural networks are biologically inspired ~\emph{Artificial Neural Networks} (ANN), which are more power-efficient, but the discrete nature of the spikes makes it difficult to train an SNN. Thus, much work has been done to convert the topology of a trained network from ANN or CNN to SNN while copying the trained weights and adjusting the spiking threshold potential in SNN. In one such work~\cite{deng2021optimal}, Deng et al. analyze the ANN to SNN conversion error and propose a method to transfer weights with no loss of accuracy. The authors modify the ReLU function on the source ANN by thresholding the maximum activation and reducing the simulation time to 1/10th of a typical SNN simulation time. Followed by a shifting operation to balance the output frequency is termed as the threshold balancing mechanism~\cite{sengupta2019going}. 

The ANN-SNN conversion techniques have managed to achieve deeper architectures but most of these techniques are based on the assumption that spike patterns of SNN a time-sampled ANN. Thus, the temporal properties of SNN are ignored. Some training algorithms like ~\emph{Backpropagation Through Time} (BPPT) could solve this by training through time but it introduces vanishing gradients. Samadzadeh et al. in~\cite{samadzadeh2020convolutional} instead initialize the weights with non-spiking training data. For the first few epochs, the SNN is trained like an ANN where the output activation function is changed to leaky ReLU. After which the activation function is modeled as a step function with the LIF neuron. By utilizing the skip connections in~\emph{ResNet} architecture a~\emph{Spiking ResNet} (STS-ResNet) architecture is proposed for testing the conversion accuracy and temporal feature extraction on~\emph{CIFAR10-DVS},~\emph{NMNIST} and~\emph{DVS-Gesture}.

The spatial and temporal properties of the SNN can be preserved to a greater extent with the use of recurrent convolution SNN. A sampling window in recurrent architecture captures the temporal correlations in event-based sequences. In~\cite{xing2020new}, Xing et al. use a supervised~\emph{Spike Layer Error Reassignment} (SLAYER) training mechanism for ANN-SNN conversion. The SCRNN architecture is a combination of single SCRNN cells that process input sequences separately to maintain the temporal dimension. The spatial complexity is handled with decomposed input processed in cells at every time step. Each SCRNN cell accepts an input feature map, a fused feature map of previous states, and an output feature map of the current input. 
Thus, the network was made recurrent and evaluated on the \emph{Dynamic Vision Sensor} (DVS) gesture dataset, achieving an accuracy of $96.59\%$ for a 10-class gesture recognition task.

The elimination of all matrix multiplications in SNN makes them more energy-efficient than CNN. In~\cite{article}, Wu et al. develop a unified framework that supports weight normalization, threshold balancing and an~\emph{Explicit Current Control} (ECC) method that controls the number of spikes passed to the SNN and reduces the residual membrane potential. The residual current in the neurons causes accuracy loss for shorter inference. The ECC also enables the conversion of~\emph{Batch Normalization} (BN) layer that many of the recent works fail to include. The role of the BN layer is to normalize the previous layer's output, which accelerates the CNN's convergence. This is implemented in the conversion by a numerical constant dependent on the training platform that updates the weight and bias to a normalized version of the same.

\subsection{Neuron models on FPGA}
A biological neuron can be implemented in either of the neuron models as discussed in~\autoref{sec:background}. In~\cite{kumar2016design}, Juneeth et al. develop a spiking neuron model of \emph{Hodgkin-Huxley} on an FPGA. The hardware architecture is based on a series of adder and multiplier blocks to implement the first-order differential equations in Verilog on~\emph{Xilinx Virtex-5}. Each of the neuron parameters is represented in a fixed point with chosen bit lengths. However, the physical justification for the bit optimization strategies is inconsistent. The use of multiplier blocks increases the number of LUTs and eventually, power consumption.

On the contrary, in~\cite{shama2020fpga}, Farzin et al. propose a multiplier-less \emph{Hodgkin-Huxley} model. The model approximates hyperbolic functions as piece-wise linear terms implement all multiply operations as logical shifts and adds. This is made possible with equations modified to power-2-based functions. Considering the high accuracy of a HH neuron, a multiplier-less design furthermore reduces the operational cost and increases the frequency. 

Two simplified two-dimensional versions of the HH neuron model are the ~\emph{FitzHugh-Nagumo} and~\emph{Morris-Lecar} neuron models~\cite{gerstner2002spiking}. In~\cite{nouri2015digital}, Nouri et al. investigate the~\emph{FitzHugh-Nagumo}  neuron model in terms of its digital implementation feasibility and computation overhead. The model is a cubical approximation with no auxiliary reset equations. The area and power numbers of the proposed design with \emph{Euler} discritization are lower than a HH model apparatus. In~\cite{mellal2021flexible}, Mellal et al. study the behavior of the Morris-Lecar neuronal model on a~\emph{Xilinx Zynq UltraScale+} board. The simulated and hardware results of the original and discrete implementations have a high correlation. 

One of the basic building blocks of the human body is the~\emph{Central Nervous System} comprised of neurons, synapses, and glial cells. In~\cite{alkabaa2022investigation}, Abdulaziz et al. evaluate a model based on its complexity and hardware resources required to realize it on an FPGA. The two-neuron coupled~\emph{Izhikevich} model while satisfying the above parameters regenerates all the different dynamics of the human brain. The authors use a LUT-based approach to the quadratic equations instead of an approximation technique to replace the non-linear terms. As the size of the LUT increases to 1000 points the model becomes closer to the mathematical equations.

Another unique technique involving the \emph{Izhikevich} neuron model is discussed in~\cite{karaca2021extensive}. According to the authors, the \emph{Izhikevich} neuron is practically more suitable for electrical realization due to its chaotic behavior. A modified version of the model is compared with the original coupled neuron dynamics through numerical simulations and FPGA device demonstration. The modified model has a longer processing time (2 clock cycles) but with no multiplier usage. This design characteristic has significantly reduced the utilization number while maintaining the accuracy of the model. Along with the spinal network these neuron models have been altered for use. In~\cite{niu2012multi}, Niu et al. the motoneurons in the motor nervous system are emulated with the~\emph{Izhikevich} model on an FPGA to analyze pediatric neurological diseases.

\begin{figure}[h]
        \centerline{\includegraphics[width=\columnwidth]{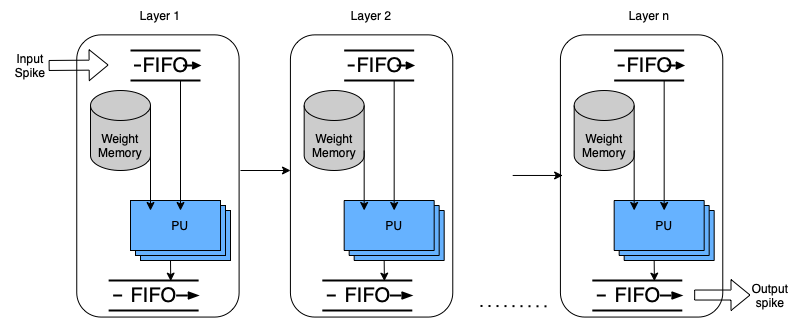}}
        \caption{Layer-by-layer architecture with LIF neuron in Processing Unit (PU).}
        \label{layerarch_fig}
    \end{figure}

In~\cite{panchapakesan2021syncnn}, Panchapakesan et al. proposed a novel synchronous SNN execution divided into layers for each input, hidden, and output layers. The input is executed layer-by-layer as shown in~\autoref{layerarch_fig} synchronously to avoid frequent off-chip memory communication. Further, the internal parameters like membrane potential can be immediately used, reducing the on-chip resources utilized. Each layer is composed of modules that implement the functionality of integrating and fire neuron structure as discussed in~\autoref{sec:background}. Such layer network depiction can also be observed in~\cite{khodamoradi2021s2n2}. Khodamoradi et al. propose a streaming SNN accelerator architecture that utilizes the concept of layer modules for generating output spikes. However, this architecture is made more specific to edge devices with sparse input data. Highly sparse events can still utilize significant memory to store null information. This is mitigated with the use of binary tensors as input buffers.

The hardware layer network is also observed in~\cite{fang2020encoding} where Fang focuses their work to retain the temporal information in a sequence of spikes in the architecture. The leaky integrate and fire neuron-based hardware avails a series of~\emph{Infinite Impulse Response} (IIR) filters to represent SNN. The time-encoded spikes provide the time information to interpret neurons as filters. Further, the suggested model incorporates synapse dynamics adding flexibility to the system. Each layer has a processing element that updates the neuron and synapse state variables. An optimized framework of end-to-end neuromorphic systems is described with reduced inference latency. The results compared to a CPU has better performance metrics.

Additionally, the LIF model has been holistically modified to several other forms. In~\cite{heidarpour2016cordic}, Heidarpour et al present a~\emph{CORDIC} implementation of an ~\emph{Adaptive exponential integrate and fire} (AdEx) neuron model. The primary difference to~\autoref{lif_formular} is that the potential difference integrates exponentially in the current equation. In a biological neuron, the increase in action potential with an input spike is non-linear. Thus, non-linear LIF neuron models depict the biological behavior rather closely although the implementation cost of non-linear behavior on hardware is always higher. The authors in~\cite{heidarpour2016cordic} discretized the differential equations explicitly with the Euler method and implement it on a~\emph{Xilinx Spartan 6} using simple add and shift operations only. Similarly, in~\cite{basham2012compact}, Basham et al demonstrate a~\emph{quadratic integrate and fire} (QIF) neuron model where the current equation integrates the square of potential difference. The design utilizes a fixed-point multiplier to evaluate the square of the voltage.

Lastly, in~\cite{nallathambi2020probabilistic}, Nallathambi et al instrument a probabilistic integrate and fire neuron model. In a deterministic approach when a neuron i spikes, the applied weight of the postsynaptic neuron is $w_{ij}$.For a set of $N_i$ spikes with a probability of $p_{ij}$ the new weight updates to $N_i \times p_{ij} \times w_{ij}$.In stochastic processing, the number of memory fetches can be reduced significantly. By sorting the synaptic weights the authors report a reduction of $90\%$ off-chip memory accesses.

\subsection{Optimization techniques}
\emph{Spiking Neural Networks} can be divided in two different categories of optimization strategies. One, to make more biologically plausible neural models with better plasticity and second, to devise tuning algorithms that can transcend performance. More plasticity is accomplished with a trainable spiking network. Backpropagation is one such learning mechanism that has shown significant accuracy in ANN but it is not realizable for a biological spiking network. Thus, much work has been done to create the equivalent spiking backpropagation. 

\subsubsection{Training SNN}
Training deep spiking Neural networks have shown a great requirement for optimized gradient-based approaches. One of the reasons for facing difficulty in achieving deeper DNN is its complex~\emph{Spatio-Temporal Dynamics}. \emph{Converting ANN to SNN} and \emph{Backpropagation with Surrogate Gradient} are two ways to get deep SNN. Converted SNNs need a longer training time for similar precision as ANN. Most popularly,~\emph{Backpropagation Through Time} (BPTT) is implemented by unfolding the gradient over a simulation time. In~\cite{fang2021deep}, Fang et al propose~\emph{Spike-Element-wise} (SEW)~\emph{ResNet} to resolve exploding gradient with identity mapping. In~\emph{Spiking ResNet}, the adaptation of residual learning from~\emph{ResNet}, the ReLU activation block is replaced by spiking neurons which is incapable to replicate the identity mapping for all neuron models. However, in~\emph{SEW ResNet} element-wise function "g" facilitates identity mapping. When "g" is chosen to be an ADD function it avoids the infinite outputs problem by restricting the output. 

Apart from a purely algorithmic perspective of improvement in spiking realizable Backpropagation, some work has been done by Shrestha et al to develop a biologically plausible algorithm that can fit the constraints of neuromorphic hardware. In~\cite{shrestha2021hardware}, the authors have shown in-hardware supervised learning demonstrated on \emph{Intel's Loihi} chip. The supervised learning is based on~\emph{Error modulated Spike-Timing Depnedent Plasticity} (EMSTDP) with~\emph{Direct Feedback Alignment} (DFA) that reduces the number of neuron updates in the feedback path. The hardware realization of the algorithm is designed by creating two replicas of the same neuron with each of them functional in feed-forward and feedback paths respectively.

Further in~\cite{heidarpur2019cordic}, Heidarpur et al propose on-FPGA online STDP. Based on~\emph{CORDIC} an iterative algorithm all hyperbolic and exponential functions can be implemented with shift and addition operations. A~\emph{Izhikevich} neuron behaviour was replicated with qualitative error analysis between the~\emph{CORDIC} model compared to the original model of operation to achieve maximum precision.The proposed algorithm is tested for a two-layer network of 21 neurons with one output neuron.To demonstrate STDP learning on~\emph{Xilinx Spartan-6} FPGA, all the exponential terms were approximated with a negative power of 2 terms which can be synthesized in fast and low-cost hardware. 

In STDP the weight update depends on the timing difference between the presynaptic and postsynaptic neuron pair. On the other hand,~\emph{Triplet STDP} (TSTDP) considers three consecutive spikes, one presynaptic and two post synaptic spikes for potentiation and depression respectively. In~\cite{gomar2018digital}, Gomar et al could reproduce the learning curve by approximating the two variable TSTDP equations to one variable piecewise linear term where exponentials are converted to base-2 functions. The discrete values of the lines are stored in LUT memory whose size is optimized as per the design. The design is driven through a~\emph{Finite State Machine} (FSM) of 4 states with 2 states representing the Learning unit of the system. Compared to the LUT model this model is less accurate but consumes less than $1\%$ on-chip resources.

So far we have discussed~\emph{Hebbian} learning applications, however, in~\cite{liu2019enabling}, Liu et al explored non-Hebbian on-chip learning IP on recurrent SNN. The~\emph{Liquid State Machine} (LSM) based recurrent SNN consists of a reservoir which maps the input pattern to a multi-dimensional response in the reservoir. The output of the reservoir is then passed to the readout layer. The focus of this work is to develop learning rules with~\emph{SpiKL-IP} algorithm based on neural plasticity.~\emph{SpiKL} is a~\emph{Intrinsic Plasticity} (IP) rule whose key idea is to maximize the input to output information transfer. The algorithm is modified to reduce the complexity of hardware implementation with LUTs and series expansion techniques. Moreover, the LIF neuron model is discretized to surpass differential equations. Compared to baseline LSM the current model has outperformed by $8\%$ for speech recognition~\emph{TIMIT} dataset. Although a hardware resource utilization trade-off with extra energy overhead could be observed. 

\subsubsection{Tuning SNN}
Since SNN architectures are resource-constrained, compression methods like pruning are an essential optimization to tune the performance. Nonetheless, most of the pruning approaches are ANN-specific and not compatible with SNN. This factor lingers the performance of a pruned SNN. In~\cite{chen2021pruning}, Chen et al formulate a learning algorithm of connectivity and weight that defines gradient as a different parameter called~\emph{Gradient Rewiring} (Grad R). The combined learning and pruning algorithm change the synaptic connections based on the synaptic gradient parameter introduced. The algorithm is evaluated on~\emph{MNIST} and~\emph{CIFAR-10} datasets with a maximum of $73\%$ connectivity. Although the techniques discussed in this section are the new frontiers of SNN optimization innovation, most of them have not made their way to a physical hardware implementation.



\subsection{Technique Discussion}
The design and implementation of deep learning algorithms, particularly SNN hardware implementation, has been a hot topic during the previous decade \cite{mittal2020survey}. SNN is hungry for computing power. Because of the rapid improvements in integrated circuit technology, it is now feasible to build high-performance chips at a low cost and with efficient energy consumption. This allowed for the deployment of inference at the edge. Along with, SNN can be executed on any device with adequate capability. There are some examples: i) CPUs, ii) GPUs, iii) FPGAs, iv) ASICs. Diverse technologies have a different throughput, performance, design flexibility, and power efficiency.

Firstly, we partially consider traditional computing platforms to provide a comparison overview of the latest SNN accelerators targeting low-power and high-performance. As depicted in~\autoref{table:table_1}, after developing circuits for edge computing, energy and performance efficiency are the most important criteria to consider. CPUs \cite{fang2020encoding}\cite{ju2020fpga} are the most adaptable gadgets, and as such, have the lowest performance and energy efficiency. 

Furthermore, ASICs \cite{park20197, 8008536, kim2015640m, chen20184096, cho20192048} provide the highest performance and energy efficiency, but with very little flexibility attainable primarily by additional hardware logic. It has reached a peak performance of $100K img/sec$ and $7.89TOPS/W$ mean that are very energy efficient. ASIC circuits are specifically built for neural networks to improve performance and power efficiency.

Modern GPUs incorporate single-precision computation modules capable of performing several half-precision floating-point calculations. Thus, they seem to be the perfect basis for both learning and prediction. However, due to their multi-core structure, GPUs present a challenge with high power usage. Studies \cite{fang2020encoding, ju2020fpga} have shown power consumption figures markedly higher than those for FPGAs.

FPGAs combine hardware and software (processing) in a single device, where software ensures programmability and hardware is utilized to implement specialized accelerators. GPUs have the higher performance among programmable or configurable devices and are less expensive than FPGAs.

Researchers have demonstrated that FPGAs may efficiently implement Spiking Neural Networks (SNN) in both hardware and software/hardware contexts \cite{yang2018real, pani2017fpga, han2020hardware, liu2022fpga, ju2020fpga, fang2020encoding, ma2017darwin, li2021fast, carpegna2022fpga}. They showed that their FPGA-based design had the highest performance of 2124 frame/sec and the lowest power of $0.40W$. This corresponds to an energy efficiency of $16.80mOPs/W$. The same platform may then be upgraded with an SNN without any board modifications just by reconfiguring the device.

\begin{figure}[!htbp]
        \centerline{\includegraphics[width=\columnwidth]{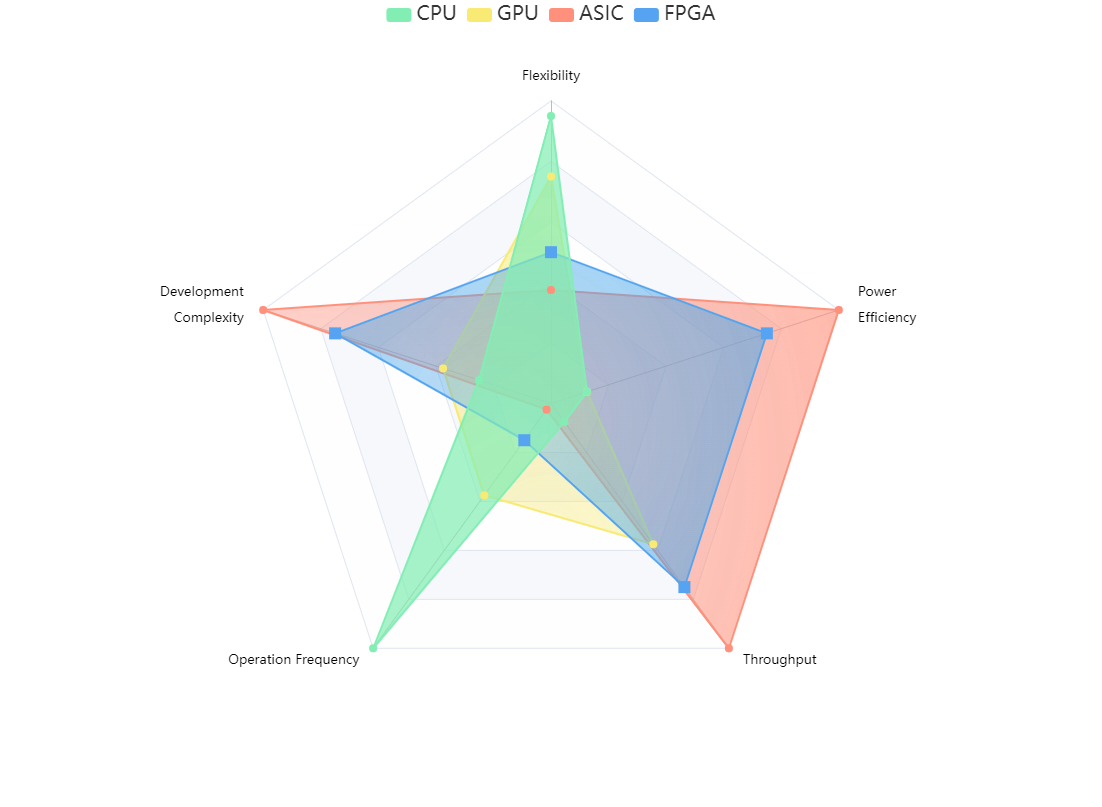}}
        \caption{SNN on Different Heterogeneous Devices}
        \label{Chart1}
    \end{figure}          

\begin{table*}[ht]
\renewcommand{\arraystretch}{1.2}
	\setlength{\tabcolsep}{3pt}
\centering
\caption{SNN on different heterogeneous devices}
\label{table:table_1}
\resizebox{\linewidth}{!}{
\begin{tabular}{llllllll}
\toprule
 Work           & \multicolumn{2}{l}{Platform}   & Operation Frequency &Power  & Technology & Throughput &Scaled Efficiency \\  \hline 
   
  Park et al.~\cite{park20197}&ASIC  & LP CMOS  &$20MHz$     &$23.6m\si{\watt}$  &$65nm$  &$100Kimg/sec$ &$7.89TOPS/W$\\
  Buhler et al.~\cite{8008536}&ASIC  & CMOS  &$250MHz$     &$87m\si{\watt}$  &$40nm$  &$1778M pixel/sec$ &$3.43TOPS/W$\\
  Kim et al.~\cite{kim2015640m}&ASIC  & CMOS  &$40MHz$     &$3.65m\si{\watt}$  &$65nm$  &$640M pixel/sec$ &$0.175TOPS/W$\\
  Chen et al.~\cite{chen20184096}&ASIC  &FinFET CMOS  &$110MHz$     &$46.6 \si{\watt}$  &$10nm$  &- &$0.110TOPS/W$\\
  Cho et al.~\cite{cho20192048}&ASIC  &CMOS  &$105MHz$     &$94 m\si{\watt} $&$40nm$  &$440M pixel/sec $&$0.083TOPS/W$\\\hline
  Yang et al.~\cite{yang2018real}&FPGA  &Stratix-III &$100MHz$  &$10.5\si{\watt}$  & $65nm$  &-&- \\ 
   Pani et al.~\cite{pani2017fpga}&FPGA  &Zynq-7  &$100MHz$     &$8.5\si{\watt}$  &$40nm$  &- &$0.290GOPS/W$\\
  Han et al.~\cite{han2020hardware}&FPGA  &Zynq-7  &$200MHz$&$0.40\si{\watt}$  &$28nm$  & $161 frame/sec$&$16.80GOPS/W$\\

  Liu et al.~\cite{liu2022fpga}&FPGA  &Kintex-7  &$200MHz$&$0.53\si{\watt} $ &$28nm$  &$208
frame/sec$ &$27.85GOPS/W$\\

Ma et al.~\cite{ma2017darwin}&FPGA  &Spartan-6  &$75MHz$&$6.3m\si{\watt} $ &$45nm$  &$208
frame/sec$ &$27.85GOPS/W$\\
 Li et al.~\cite{li2021fast}&FPGA  &Virtex-7  &$100MHz$&$0.50\si{\watt} $&$28nm$  &$317
image/sec$ &$2.08GOPS/W$\\
Carpegna et al.~\cite{carpegna2022fpga}&FPGA  &Artix-7  &$100MHz$&- &$28nm$  &$465
image/sec$ &-\\\hline

\multirow{3}{*}{}  &FPGA&ZCU-102 &$150MHz$ &$4.6\si{\watt}$  &$16nm$  &$164 frame/sec  $&$4.66GOPS/W$\\ 
                  Ju et al.~\cite{ju2020fpga} &CPU&İ7-6700k &$4.0GHz$ &$54\si{\watt}$  &$14nm  $&$4 frame/sec  $&$1.03GOPS/W $ \\
                   &GPU&GTX 1080 &$1.6 GHz$ &$100\si{\watt}$  &$16nm$  &$162 frame/sec$  &$0.061GOPS/W $ \\\hline
\multirow{3}{*}{}  &FPGA&ZCU-102 &$125MHz$&4.5\si{\watt}  &$16nm$  & $2124 image/sec $&$6.92GOPS/W$\\ 
Fang et al.~\cite{fang2020encoding}&CPU&i9-9900K &$3.7GHz$ &$58.1\si{\watt} $ &$14nm$  &$100 frame/sec  $&$2.18GOPS/W $ \\ 
 &GPU&AGX Xavier &$1.3GHz $&$14\si{\watt} $ &$12nm$  &$211 frame/sec$  &$1.67GOPS/W$\\ \bottomrule 
\end{tabular}}
 
\end{table*}

\begin{table*}[!htbp]
\renewcommand{\arraystretch}{1.2}
	\setlength{\tabcolsep}{3pt}
\centering
\caption{Different SNN models on FPGA}
\label{table:table_2}

\resizebox{0.8\linewidth}{!}{
\begin{tabular}{lllllll}
\toprule
 Work        & Model &Time Resolution  & Neurons / Core & Synapses / Neurons & \multicolumn{2}{c}{\begin{tabular}[c]{@{}l@{}}  Resources   \\({FF, LUTs})\end{tabular}} \\ \hline 
 Upegui et al.~\cite{upegui2005fpga}&Custom  &$1.0ms$  &30  &30  &100, &100    \\ 
 Pearson et al.~\cite{pearson2007implementing}&LIF  &$0.5ms$  &1120  &$\approx$912/112  & - & -    \\ 
 Cassidy et al.~\cite{cassidy2007fpga}&LIF&\SI{320.0}{\nano\second}  &51  &128  &146, &230    \\ 
 Han et al.~\cite{han2020hardware}&LIF  &$6.7ms$  &16.384  &$\approx$1.025x$10^3$  &5381, &7309  \\ 
 Thomas et al.~\cite{thomas2009fpga}&IZ  &10.0 $\upmu$s   &1024  &1024  &39, &19   \\ 
 Ambroise et al.~\cite{ambroise2013real}&IZ  &$1.0ms$  &117  &117  &8, &17    \\ 
 Cheung et al.~\cite{cheung2016neuroflow}&IZ  &$1.0ms$  &98.000  &1.000 - 10.000  &- &-   \\ 
 Pani et al.~\cite{pani2017fpga}&IZ  &$0.1ms$  &1440  &1.440  &37, &39    \\ 
 Gupta et al.~\cite{gupta2020fpga}&Simplified LIF  &$1.0ms$  &800  & 12.544  &29, &70   \\
 Li et al~\cite{li2021fast}&LIF  &-  &1094  & 162  &- &-  \\
 Asgari et al~\cite{asgari2020low}&LIF  &45.6 $\upmu s$  &16  & 7  &5090, &34648   \\
 Fang H et al~\cite{fang2018scalable}&LIF  &$1.0ms$  &984  & 1.59  &-, &57600   \\
 Ma et al.~\cite{ma2017darwin}&LIF  &$160s$  &1794  & 360  &4705, &11489   \\
 Carpegna et al~\cite{carpegna2022fpga}&LIF  &$21.5ms$  &1384  & 226  &26853 , &29145   \\
 Liu et al.~\cite{liu2022fpga}&LIF  &$7.5ms$  &16.384 &$\approx$1.025x$10^3$  &30417, &46371\\\bottomrule  
 
\end{tabular}}

\end{table*}

In this section, key ideas for SNN implementation in equipment are shown, along with a comparison to different techniques in~\autoref{table:table_2}. The table combines the most important information received from the available sources from among the numerous structures specified from now on. Without being exhaustive, this chart allows for a quick representation of the show status of the advanced systems.

This work \cite{cheung2016neuroflow}, the usage of the \emph{NeuroFlow} which FPGA architecture for SNN that was suggested approach aims for a real-time execution time of $0.1ms$. \emph{NeuroFlow} may be further reduced by adjusting the size of the arrangement or guessing on the network within this architecture. Moreover, the huge quantity of simulated neurons made available by \emph{NeuroFlow} may be achieved using six FPGAs in a toroidal network design. The configuration restricts the number of synapses to a range of 1,000–10,000, where the chance of connection is based on a Gaussian probability of synaptic disconnection, with a standard deviation varying from 32 to 512. In this spectrum, a direct comparison might be challenging and potentially unfair to other SNN models. Another approach in this direction has been implemented by Han~\cite{han2020hardware}, which highlights crossover upgrade algorithms that integrate the points of interest of current algorithms to improve equipment planning and execution. Thus, this design, capacitively supports 16 384 neurons and 16.8 million synapses but uses fewer hardware resources and has a very low power consumption ($0.477 \si{\watt}$) and the performance for processing neuron activation events is $6.72ms$. 
For a modest to large scale SNNs on sophisticated equipment should be computationally simple and simultaneously capable of communicating to the large range of termination patterns displayed by various organic neurons. For this reason, several designs employ simpler custom neural models~\cite{upegui2005fpga}, and much more with the LIF model \cite{cassidy2007fpga, pearson2007implementing}.
Spikes collection is performed at $32 bits$ to maintain as much accuracy as conceivable. In \cite{ambroise2013real}, the authors demonstrate a comparable technique capable of simulating up to 167 neurons. In comparison to the one suggested, due to increased information exactness and synaptic current preparation which is more sophisticated to a certain degree such a task employs a greater number of assets.
The FPGA architectures by~\cite{thomas2009fpga} and~\cite{pani2017fpga} simulate a fully-connected network of 1024 neurons, 1440 neurons, based on the biologically plausible \emph{Izhikevich} spiking model. The reported time resolutions are 10.0 $\upmu s$ and $0.1ms$. These implementations have low latency, despite the fact that simplified resource utilization is employed for it. An alternate notable design is that by Gupta et al.~\cite{gupta2020fpga}, which offers a simulation of a simpler and more computationally efficient model using FPGA infrastructure. This was engineered to leverage the sparsity of the network and generate each time unit corresponding to network activities. A network comprising 784 input and 16 output neurons, along with 12,544 synapses, was realized on hardware, thereby facilitating the creation of a hardware accelerator with minimal resource consumption.

In another study by Liu et al.~\cite{liu2022fpga}, a neuron computing module was used, designed to replicate both \emph{LIF} and \emph{Izhikevich} neurons using the concurrent spike caching and scheduling strategy while simulating 16,384 neurons and 16.8 million synapses. They structured two distinct three-layer SNN networks applied for recognition tasks on the suggested platform. The employment of Slice LUTs, Slice Registers, and DSP of the SNN acceleration unit stands at $6.0\%$, $2.5\%$, and $7.6\%$ respectively.

It is noteworthy that their design exhibits the highest occupancy ratio in BRAM. As a consequence, the parameters of the LIF and IZH neurons, along with the event buffer and synaptic delays, are produced using the on-chip storage resource BRAM. Hence, the BRAM is significantly employed, leading to increased power usage. Ma et al.~\cite{ma2017darwin} established an embedded auxiliary processor to hasten the inference mechanism of SNNs. By time-sharing the physical neuron components, and developing a reconfigurable memory subsystem, the design realizes high hardware efficiency, fitting for resource-limited embedded applications. It also permits a customizable quantity of neurons, synapses, and synaptic delays, yielding significant flexibility.

Moreover, work by Fang H et al~\cite{fang2018scalable}, Asgari et al~\cite{asgari2020low}, and Li et al~\cite{li2021fast} proposed to utilize their design on the identical neuron model~\emph{(LIF)} and targeting the same dataset~\emph{(MNIST)}. Recently, Carpegna et al~\cite{carpegna2022fpga} presented Spiker, which employs a clock-driven neuron design wherein the membrane potential is modified at each clock cycle absent any spikes. Nevertheless, inputs are only processed when there's at least one spike present at the input of a layer. This approach uses more power than purely event-driven designs but allows for the utilization of fewer hardware resources. The FPGA usage is of 1384 neurons and 313600 synapses are  around $55\%$ for the LUTs, and $25\%$ for the FFs. Considering the number of instantiated neurons, this is a major finding.

\section{Applications}
\subsection{Image and Audio Processing}
In~\cite{cerezuela2016sound}, authors presented an application of animal behavior recognition using an SNN-based sound recognition system associated with animal movements. The spiking neural network was built on an FPGA device, and the neuromorphic auditory system employed in this study creates a representation similar to the spike outputs of the biological cochlea. Even when the sound was accompanied by the white noise of the same strength, the detection system based on SNN achieved an accuracy of over 91\%.

\begin{figure}[h]
        \centerline{\includegraphics[width=\columnwidth]{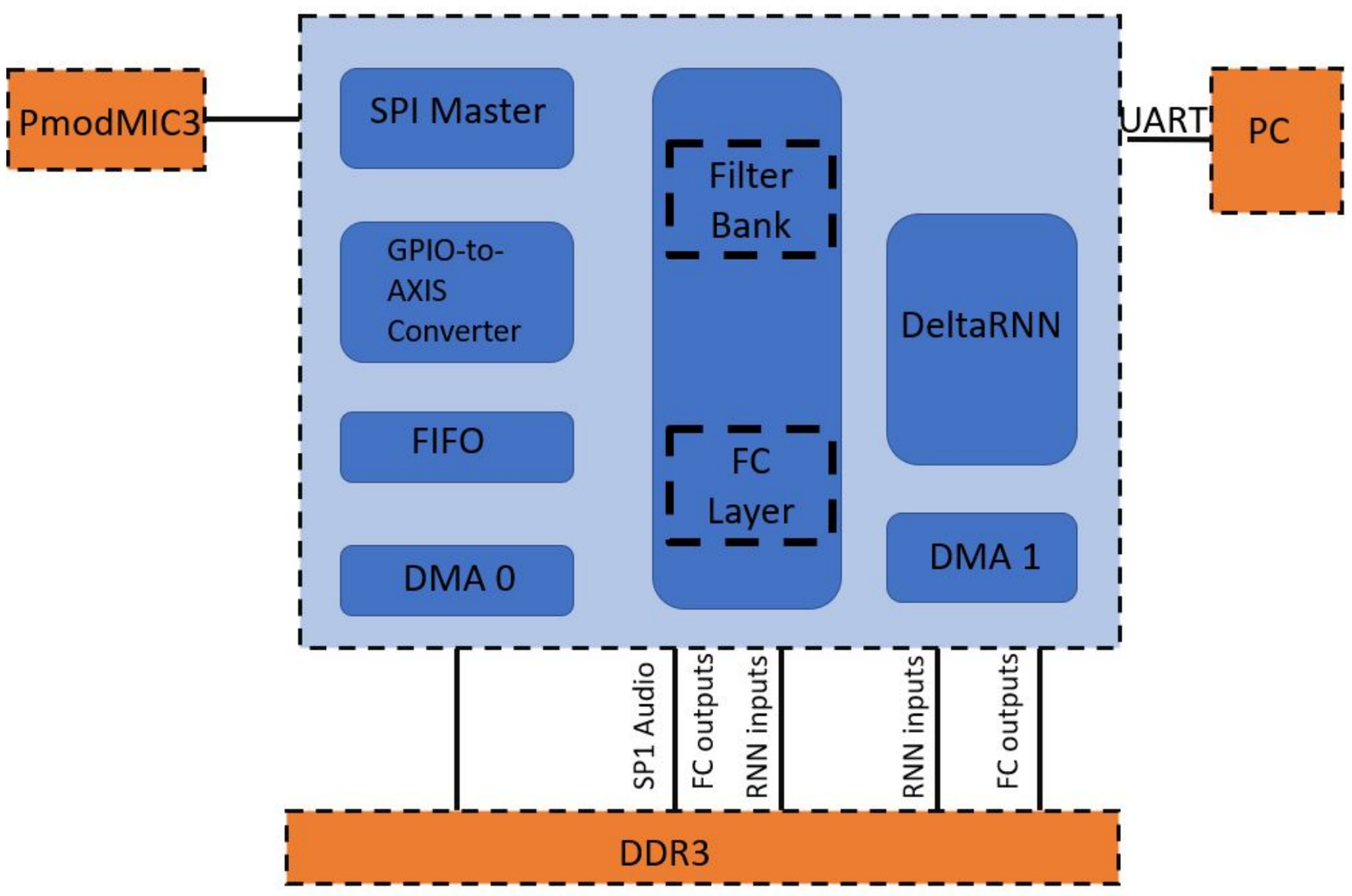}}
        \caption{System architecture of microphone}
        \centering
        \label{Overview of ELAS}
    \end{figure}

In~\cite{gao2019real}, the authors proposed a continuous voice recognition system that employs a~\emph{Delta Recurrent Neural Network} (DeltaRNN) on a~\emph{Xilinx Zynq-7100} FPGA to provide low latency~\emph{Recurrent Neural Network} (RNN) processing. This system is suitable for~\emph{Internet of Thing} (IoT) applications since FPGA uses just $70mW$ and can process each feature frame in microseconds.

Edge detection is a prominent machine learning job that necessitates the use of a large number of neurons as well as software simulations. As a result, scalability suffers, and computational time increases. This research ~\cite{glackin2009emulating} proposed a scalable FPGA implementation strategy with several innovations to shorten calculation times. Within this scenario, given spatial constraints, the time-division multiplexing structure proposes a balance between acceleration performance and the dimensions of SNN simulations. Furthermore, a study by Louis-Charles Caron et al. ~\cite{caron2011fpga} accomplished image segmentation and monophonic audio source separation utilizing \emph{Oscillatory Dynamic Link Matcher} (ODLM) protocols for motif recognition.

The authors in~\cite{bhuiyan2009character} developed a character identification model for multicore architectures relying on two SNN models. The selection of the \emph{Izhikevich} and \emph{Hodgkin-Huxley} neuron models over the integrate and fire models were motivated by their biological accuracy. All 48 visuals in the learning datasets were accurately distinguished within $14ms$ and 3.75ms, respectively. Moreover, the authors of~\cite{nuno2012hardware} employed SNN for image clustering using FPGA. The \emph{Gaussian Receptive Field} (GRF) is the most commonly used method for encoding data, with Hebbian learning being used for neural models. Not only did ~\cite{nuno2012hardware} use the \emph{Receptive Field} (RF) for data encoding, but ~\cite{iakymchuk2014hardware} also applied the T.Iakymchuk model. To examine an image in its spatial domain form, the authors also utilized the Gabor filter, a bandpass filter.

A biologically driven, rotationally invariant visual identification system, executed using a pixel camera on FPGA, was deployed in~\cite{sofatzis2014rotationally}. The structure merged the \emph{Ripple Pond Network} (RPN), a neural network proficient in fundamental 2D to 1D image conversion, rotationally invariant \emph{Temporal Patterns} (TPs), and the \emph{Synaptic Kernel Adaptation Network} (SKAN). By using the event-driven \emph{Spike-timing dependent plasticity} (STDP) rule, Zaibo Kuang and Jiang Wang ~\cite{kuang2019digital} trained the MNIST dataset for digital categorization on a \emph{Stratix-3} FPGA processor with $93\%$ precision. As a result, the cost-effective network demonstrated great efficiency and can readily be scaled to a larger size.

\subsection{Biomedical Applications}

The manuscript~\cite{lu2018spwa} presents a freely available FPGA-centric emulation environment for delving into the neuromorphic computation. The authors implemented the~\emph{MNIST}~\cite{lecun1998mnist} dataset and\emph{Vector-Matrix Multiplication} (VMM), comparing their accuracy with analogous architectures generated by IBM's Compass simulation infrastructure. Considering the substantial computational intricacy of spiking neural networks, it's arduous to actualize them on hardware necessitating proficiency. Thus, QingXiang Wu~\cite{wu2015development} unveiled a straightforward, effective, and swift strategy to instantiate SNN employing a customized toolkit. Thus, participants in this investigation seamlessly engineered and modeled various spiking neural networks on FPGA, enhancing the execution speed.

In a different study, Junxiu Liu~\cite{liu2018bio} proposed a novel biologically-derived gas identification technique with minimal hardware footprints. This method utilized spiking neural pathways and neurons to recognize the abnormal frequency of the input spike generation by adjusting the discharge likelihood of the inhibitory neural linkage under diverse input situations.

\begin{figure}[!htbp]
        \centerline{\includegraphics[width=\columnwidth]{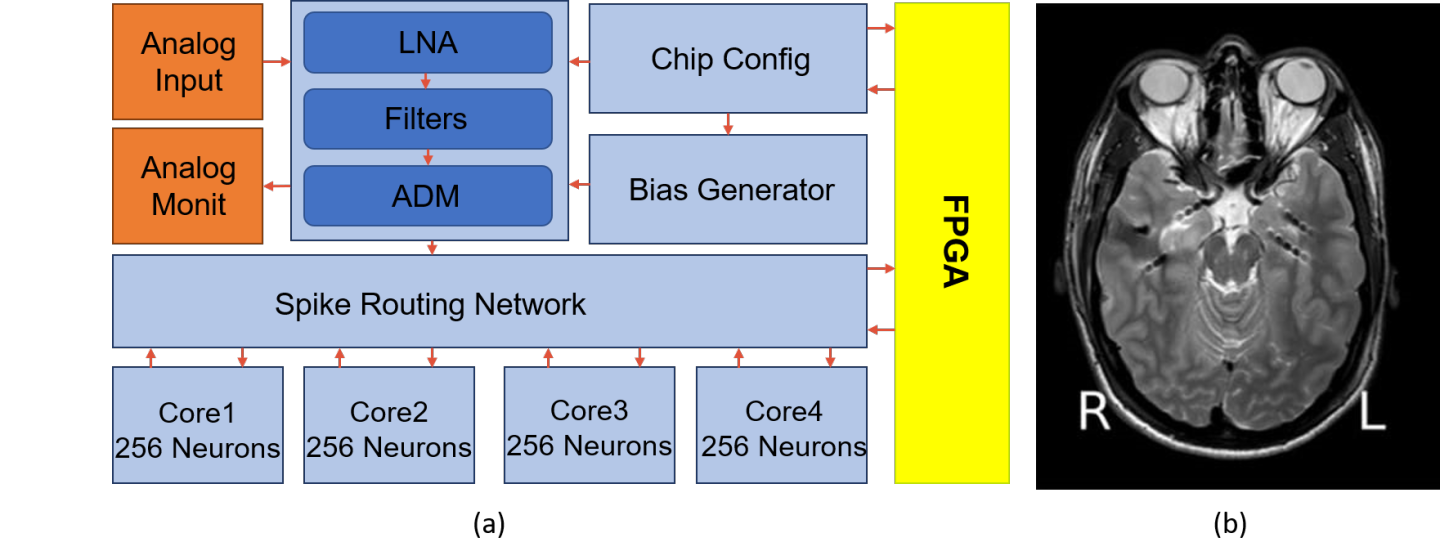}}
        \caption{ Block diagram of the neuromorphic platform (a) and~\emph{Functional magnetic resonance imaging} (fMRI) brain image (b)}
        \label{Block diagram of the neuromorphic platform}
    \end{figure}

In~\cite{guerrero2007attractor}, a neuromorphic system on a~\emph{Virtex-2 pro} FPGA was shown how neuronal elements from this construction kit can be deployed in parallel. The neuromorphic system mirroring the known architecture of the early olfactory pathway when coupled spontaneously produces rhythmic, limit-cycle dynamics resembling biology. Thus, The findings of this work show how these attractors may be read from the network to recognize previously experienced learned fragrances without confusion.

The authors~\cite{sharifshazileh2021electronic} present a neuromorphic system that combines a neural recording headstage with an SNN processing core on the same die for processing~\emph{iElectroencephalography} (iEEG), and demonstrate how it can reliably detect~\emph{High-Frequency Oscillations} (HFO), achieving state-of-the-art accuracy, sensitivity, and specificity. This is the first feasibility research aimed at finding significant characteristics in iEEG in real-time utilizing mixed-signal neuromorphic computing Technologies on a custom board~\emph{XEM7360} FPGA.

In~\cite{rana2022fpga}, The researchers wanted to construct an artificial intelligence signal identification system in a~\emph{PYNQ-Z2} FPGA board that can detect patterns of bio-signals such as~\emph{Electrocardiography} (ECG) in battery-powered edge devices. Despite the increase in classification accuracy, deep learning models need expensive processing resources and power. Therefore, it makes deep neural network mapping a time-consuming process and deployment on wearable devices difficult to achieve. SNNs have been used to overcome these constraints.

\subsection{Control Systems}

\begin{table*}[!htbp]
\renewcommand{\arraystretch}{1.2}
	\setlength{\tabcolsep}{3pt}
\centering
\caption{Applications of Control Systems}
\label{table:table_4}
\resizebox{0.8\linewidth}{!}{
\begin{tabular}{llllllll}
\toprule
 Work   & Task & Platform & Neuron Model     & Power  &Frequency  & \begin{tabular}[c]{@{}l@{}}  Resources   \\\end{tabular} \\ \hline 
 
 Jimenez et al. \cite{jimenez2012neuro} & spike-based PID&Spartan-3 &LIF  &65  m\si{\watt}  &145 MHz    &167 slice registers    \\
 Jalilian et al. \cite{jalilian2017pulse} & PWM &Spartan-6& Izhikevich    &14  m\si{\watt}  &57 MHz     &499 slice registers    \\
 Kotlyarevskaya et al. \cite{kotlyarevskaya2018control} & Myoelectric Control &Virtex-6& LIF &-  &-   &-    \\
 Linares et al. \cite{linares2020ed} & spike-based PID &Spartan-6 & Custom &12  \si{\watt}  &140 MHz   &-    \\\bottomrule
 
\end{tabular}}
\end{table*}
In~\cite{jimenez2012neuro}, proposed a procedure that permits the usage of analog-like spike-based~\emph{Proportional–Integral–Derivative} (PID) controllers on low-cost~\emph{Virtex-3} FPGA. Spike-based PID controllers have been logically analyzed and characterized, demonstrating that they are exceptionally near to nonstop controller models unless they are actualized in a totally computerized gadget as FPGAs.

In~\cite{jalilian2017pulse}, biological neuron model-generated spiking waveforms were utilized as substitutes for sawtooth waveforms to effectuate~\emph{Pulse-Width Modulation} (PWM), a modulation strategy that produces variable-width pulses to symbolize the amplitude of an analog input signal. This study proposes a signal generator on the FPGA platform, grounded in the~\emph{Izhikevich} neuron model, prized for its simplicity and precision.

An innovative stride in the realm of prosthetics is the bio-electric upper limb prosthesis serving as an actuator's control. The authors of~\cite{kotlyarevskaya2018control} adopted the~\emph{Izhikevich} neuron model and an FPGA-oriented LIF to mimic the behavior of biological neurons. Moreover,~\emph{PWM} has surged in its acceptance as a governing technique in both analog and digital circuits.

In~\cite{linares2020ed}, a spike-oriented proportional-integral-derivative engine speed controller was modified to govern the position of the 4 joints of a lightweight and~\emph{safe physical human-robot interaction} (pHRI) robotic arm, referred to as~\emph{event-driven BioRob} (ED-BioRob). These~\emph{spiking PID} (sPID) controllers were deployed on two~\emph{Spartan-3} FPGA platforms, that is, the~\emph{address-event representation Robot} (AER-Robot). The robot furnishes address-event-representation interfacing for spiking systems and is capable of driving DC motors with Pulse Frequency Modulation signals, reflecting the motor neurons of mammals.

\subsection{Autonomous Systems}
\begin{figure}[!htbp]
        \centerline{\includegraphics[width=\columnwidth]{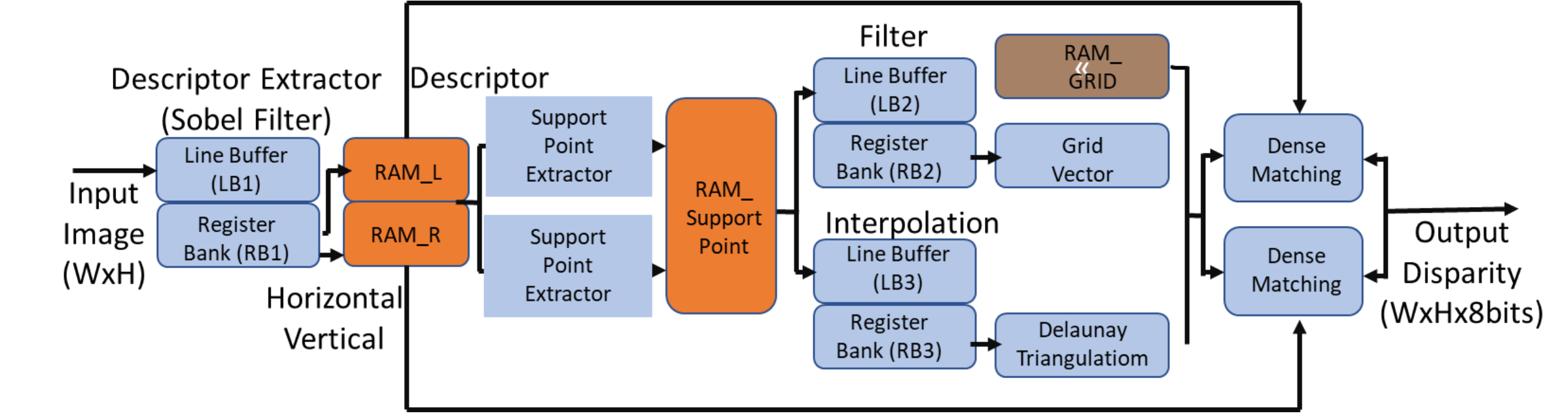}}
        \caption{Overview of ELAS}
        \centering
        \label{Overview of ELAS}
    \end{figure}
FPGAs are gaining popularity because of their reconfigurability and hardware efficiency, and have been proposed for robotic vision. The publication~\cite{gao2021ielas} introduced the~\emph{Efficient Large-scale Stereo} (ELAS) based stereo vision system that is completely implemented on FPGA and is intended for real-time and energy-efficient applications. The~\emph{ELAS} which is shown in~\autoref{Overview of ELAS} calculation is reformulated as a standard design with a back focus introduction for hardware execution.

Within the examination at~\cite{wan2021energy} the bound-together compute bottleneck of different localization frameworks is distinguished. Therefore, an~\emph{Oriented-Fast and Rotated-BRIEF} (ORB) based visual frontend design is displayed for real-time and energy-efficient localization and assessed on the FPGA platform.

Authors illustrated~\cite{murray2016microarchitecture} an end-to-end usage on a genuine, high-\emph{Degrees of Freedom} (DOF) mechanical arm, and the quickening agent was able to unravel energetic pick-and-place scenarios with a high rate of success. This sub-millisecond speed is adequate with a few limitations to empower already the infeasible automated applications, such as real-time arranging in energetic situations.

\subsection{Robotics}
Due to their handy and beneficial assistance, robots are increasingly being integrated into our society. The authors of~\cite{nhu2017cerebellum} aimed to leverage FPGA to create a cerebellum model capable of learning and adjusting conversational robot timing. The~\emph{Central Pattern Generator} (CPG), a locomotion mechanism for multi-legged robots, is an SNN application as well.

The research conducted in~\cite{mokhtar2007autonomous} applied advanced hippocampal pyramidal neuron models to construct hardware-spiking neural networks for navigation purposes. J.Parker Mitchell~\cite{mitchell2017neon} utilized a~\emph{Dynamic Adaptive Neural Network Array} (DANNA) framework for Autonomous Robotic Navigation. The DANNA framework comprises a grid of adaptable neuromorphic computing components, each capable of behaving like a neuron and connected to its adjacent counterpart in the grid.

A modular hardware execution of a spiking neural network with real-time adjustable connections is presented in~\cite{roggen2003hardware}, framed within an autonomous robot impediment evasion schema. The structure is situated in a conventional 2D matrix of cells, each operating as a spiking neuron with unique functions. Another study~\cite{johnson2017homeostatic} likewise tackled the robot obstacle avoidance dilemma. Inspired by the robustness and adaptability of biological systems, this research offered a uniquely flexible neural network model. The network observed a reduction of up to $75\%$ of the initial synaptic inputs to a cell.

In~\cite{guerra2017fpga}, a neuromorphic framework was established on a~\emph{Spartan-6} FPGA to enable locomotion for three types of robots: bipedal, quadrupedal, and hexapods. In this exploration, the researcher conceived a fast, compact, and configurable FPGA architecture that hinged on the~\emph{CPG}, the movement mechanism for multi-legged robots.

\subsection{Event vision sensors}
A fault injection experiment and fault resilience analysis for an SNN hardware accelerator were given in the study at~\cite{spyrou2022reliability}. They created a fault injection framework that builds and maps SNN faulty instances into the hardware. The framework is allowing the researchers to expedite fault injection and analyze fault criticality on~\emph{ZCU104} FPGA development board.

    \begin{figure}[!htbp]
        \centerline{\includegraphics[width=\columnwidth]{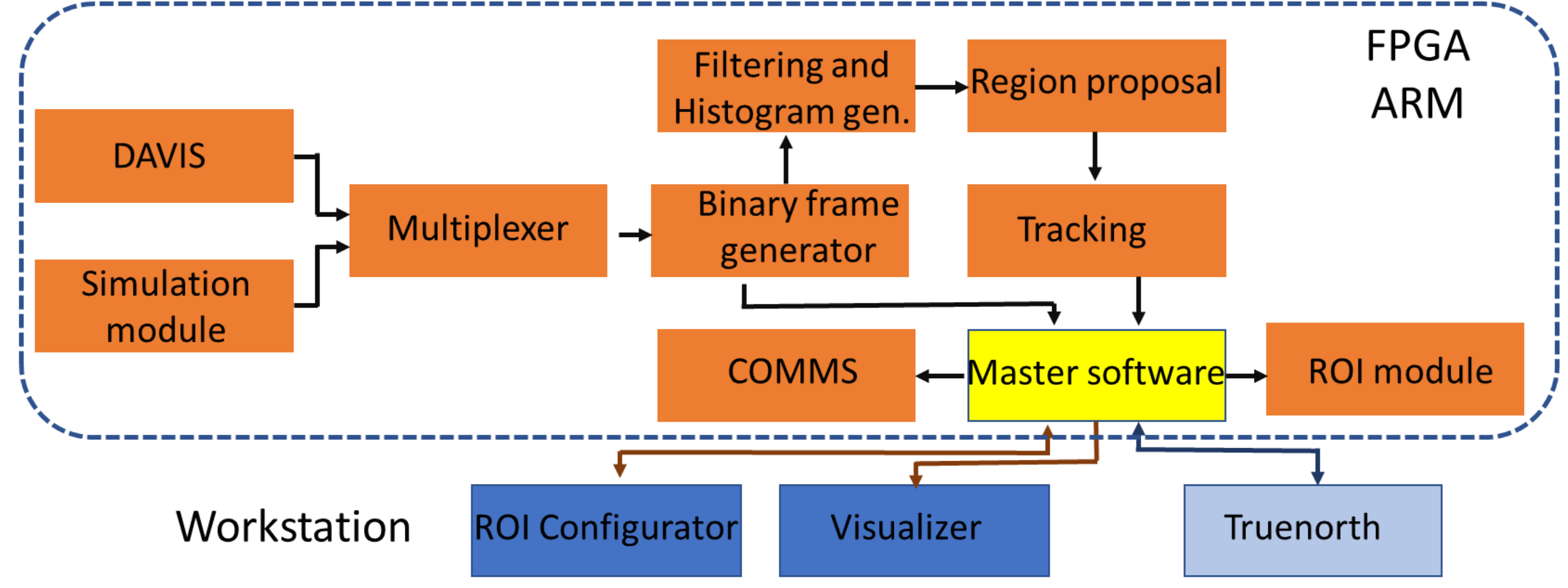}}
        \caption{System flow diagram of the neuromorphic vision sensor}
        \label{System flow diagram of the neuromorphic vision sensor}
    \end{figure}

This research described~\cite{ussa2020hybrid} one of the first end-to-end neuromorphic frameworks for real-time object tracking and categorization shown utilizing a low-power hardware implementation on~\emph{Trenz TE0720} FPGA. To take use of the low latency and asynchronous nature of~\emph{Neuromorphic Vision Sensors} (NVS) as an optional paradigm. The framework's integral operation includes gathering occurrences from the camera, handling these occurrences to isolate surveilled entities, forwarding these observed coordinates to the neuromorphic chip for sorting, and the entity detection illustration, as demonstrated in~\autoref{System flow diagram of the neuromorphic vision sensor}.

\subsection{Communication}
The authors of the work~\cite{ding2022hybrid} describe a hybrid-mode router architecture for large-scale neuromorphic simulation by merging two types of router schemes. These router schemes are proposed to allow chip-to-chip transmission of spike and non-spike data. This work is being tested on a neuromorphic platform constructed using an chip~\emph{Artix-7} FPGA.

\begin{table*}[!htbp]

\renewcommand{\arraystretch}{1.2}
	\setlength{\tabcolsep}{3pt}
\centering
\caption{Data Transmission Scheme Comparision}
\label{table:table_5}

\resizebox{0.8\linewidth}{!}{
\begin{tabular}{lllllll}
\toprule

 Platform&Technology  &Routing Method& Router Frequency& Router Latency&Peak Spike Throughput  \\\hline
 SpiNNaker~\cite{lagorce2015breaking}&ASIC(130nm)  &Multicast  &$180MHz$&$280ns$&$189Mspike/s$ \\
Loihi~\cite{davies2018loihi}&ASIC(14nm)  &Unicast  &$1.67GHz$&$6.5ns$&$160Mspike/s$\\
{Ding et al.~\cite{ding2022hybrid}}&{FPGA(28nm)}&{Multicast/Unicast}&\textbf{$200MHz$} &{$25ns$}&{$200Mspike/s$}\\\bottomrule

\end{tabular}}

\end{table*}
In~\autoref{table:table_5} displays the comparison of data transmission exists similar works. The proposed routers separate themselves from prior efforts by supporting both source-driven and destination-driven packet representations. The FPGA implemented method has a spike processing rate of $200Mspikes/s$. The research found that their algorithm, routing strategies, and router design improved communication efficiency in the specified multichip network.

\section{Trend}
Based on the above surveyed related works, we can conclude the following four potential topics in further FPGA-based SNN accelerator research:

\begin{itemize}
    \item \emph{\textbf{Approximation Computing in SNN:}} Considering the hardware resource limitation on FPGA platforms, further compressing the memory consumption of weights and parameters in SNN models is necessary. The approximation computing methods and~\emph{Posit} computing methods applied in work~\cite{approxNN}\cite{ExPAND} are the upcoming techniques that can be applied in SNN accelerators on FPGA. Besides, the application of quantization and binarization techniques on SNN accelerators can also be potential.
    
    \item \emph{\textbf{Implementation Toolchain/Framework for SNN on FPGA:}} Considering the complexity of FPGA development, as the python-based hardware generation framework shown in HLS4ML~\cite{HLS4ML}\cite{hls4ml2}\cite{fahim2021hlsml} and FINN~\cite{umuroglu2017finn}\cite{Blott2018}\cite{Rybalkin2018}, an automatic hardware design generation framework/library/toolchain, will highly benefit the implementation and deployment of SNN acceleration on FPGA platforms. These frameworks/libraries/toolchains should be able to cooperate with the training/accelerating software libraries and convert the trained SNN model to the suitable hardware design for targeting platforms.
    
    \item \emph{\textbf{Automatic Network Generation of SNN on FPGA:}} The automatic network model generation techniques, such as~\emph{Network Architecture Search} (NAS) are the upcoming topic in recent research. Work~\cite{nas4snn} explored the application of NAS on FPGA, which could also be one potential research and application idea for SNN acceleration on FPGA. 
    
    \item \emph{\textbf{Upcoming Network Models of SNN on FPGA:}} Recent work of FPGA-based SNN acceleration focus on the classic MLP, CNN, and LSTM models. Some state-of-the-art works, such as~\cite{trans4snn}\cite{zhu2022spiking}\cite{gnn4snn}, explored the application of SNN on~\emph{Transformer} and~\emph{Graph Neural Networks} (GNNs). The implementation of FPGA-based accelerators for the above networks can extend the application scenarios of related research.
\end{itemize}

\section{Conclusions}
We have provided a survey on the hardware implementation of SNN that, in particular, covers recent trends in deploying FPGAs in different applications. The field of FPGA has proven that, despite the already apparent exceptional results, various researchers have illustrated the advantages of reconfigurable hardware implementations over other heterogeneous devices. In addition, we explained how different neural networks perform on FPGAs. With this survey, we proposed to summarize state-of-the-art work and highlight critical directions of inquiry about that such as image/audio processing, robotics, autonomous systems, event vision sensor, and biomedical applications.



\bibliographystyle{IEEEtran}
\bibliography{reflibrary}

\end{document}